\documentclass[10pt]{article}
\usepackage{amsmath}
\usepackage{graphicx}
\usepackage{color}
\usepackage{orcidlink}
\usepackage{hyperref}%\usepackage[colorlinks=true,linkcolor=blue]{hyperref}
\hypersetup{colorlinks=true, linkcolor=blue, citecolor=blue, urlcolor=blue}
\usepackage{caption}
\usepackage{subcaption}
\begin{document}
\baselineskip=16pt

\begin{center}
{\large {\bf PDM relativistic quantum oscillator in Einstein-Maxwell-Lambda space-time }} 
\par\end{center}

\vspace{0.3cm}
 
\begin{center}
{\bf Faizuddin Ahmed\orcidlink{0000-0003-2196-9622}}\footnote{\textbf{faizuddinahmed15@gmail.com}}\\
\vspace{0.1cm}
{\it Department of Physics, University of Science \& Technology Meghalaya, Ri-Bhoi, 793101, India}\\
\vspace{0.2cm}
{\bf Abdelmalek Bouzenada\orcidlink{0000-0002-3363-980X}}\footnote{\textbf{abdelmalekbouzenada@gmail.com ; abdelmalek.bouzenada@univ-tebessa.dz}}\\
\vspace{0.1cm}
{\it Laboratory of theoretical and applied Physics, Echahid Cheikh Larbi Tebessi University, Algeria} 
\par\end{center}

\vspace{0.3cm}

\begin{abstract}
In this analysis, we study the relativistic dynamics of quantum oscillator fields within the context of a position-dependent mass (PDM) system in the background of a curved space-time. The chosen curved space-time is generated by a magnetic field incorporating a non-zero cosmological constant called Einstein-Maxwell-Lambda solution. To analyze PDM quantum oscillator fields, we introduce a modification into the Klein-Gordon equation by substituting the four-momentum vector $p_{\mu} \to \Big(p_{\mu}+i\,\eta\,X_{\mu}+i\,\mathcal{F}_{\mu}\Big)$, where various four-vectors are defined by $X_{\mu}=(0, r, 0, 0)$, $\mathcal{F}_{\mu}=(0, \mathcal{F}_r, 0, 0)$ with $\mathcal{F}_r=\frac{f'(r)}{4\,f(r)}$, and $\eta$ is the mass oscillator frequency. The radial wave equation for the modified Klein-Gordon equation is derived and subsequently solve for two distinct scalar multipliers: (i) $f(r)=e^{\frac{1}{2}\,\alpha\,r^2}$, and (ii) $f(r)=r^{\beta}$, where $\alpha \geq 0$, and $\beta \geq 0$. The resultant approximate energy levels and wave function for quantum oscillator fields are demonstrated to be influenced by the cosmological constant and the geometrical topology parameter which breaks the degeneracy of the energy spectrum. Furthermore, we observed noteworthy modifications in the approximate energy levels and wave function when compared to the results derived in the flat space. 
\end{abstract}

\vspace{0.1cm}

\textbf{keywords:} Einstein-Maxwell Space-times; Cosmological Constant; Relativistic Wave Equations; Position Dependent Mass System; Solutions of Wave Equation; Special Functions.

\vspace{0.1cm}


\section{Introduction}

Quantum field theory (QFT) in curved space-times \cite{M0, M1, M2, RMW, NDB, SAF} provide a framework for analyzing quantum phenomena in the presence of gravitational effects \cite{M3, M4, M5}. This theory describes quantum fields interacting with classical gravitational fields, assuming that the quantum nature of gravity is not significant in the considered regimes. In this context, gravity is modeled as a classical, curved space-time following the principles of general relativity. The resulting theory involves the propagation of quantum fields on a curved background manifold. This approach draws inspiration from the successful application of similar methods in quantum electrodynamics, where early calculations treated the electromagnetic field as a background interacting with quantized matter. Results from this approximation were found to be consistent with the full theory of quantum electrodynamics. Quantum field theory in curved space-times is particularly valuable when analyzing phenomena at scales where gravitational quantum effects are negligible, such as in the standard model of particle physics, which is applicable at scales around $10^{-19}$ meters. The theory's range of validity extends to larger scales, encompassing a wide array of interesting phenomena. One notable example is Hawking radiation \cite{M6, M7, M8}, discovered by Stephen Hawking, which predicts particle creation near black holes \cite{M9, M10, M11, M12}.

The initial success of QFT in curved space-times stemmed from discoveries like particle creation in expanding universes \cite{LP}, the prediction of black hole radiation \cite{SWH}, and the related observation of a finite temperature experienced by uniformly accelerating observers in a vacuum state \cite{WGU}. Recent applications include its use in inflationary cosmology, where quantum field fluctuations contribute to understanding early universe density fluctuations (see, for example, Ref. \cite{SW}). Beyond its applications, study of QFT on manifolds offers intrinsic value. It enhances our understanding of the structure of QFT, shedding light on aspects tied to specific space-times and their symmetries versus those that are more fundamental. Structures like the existence of a vacuum state, a well-defined particle picture, and global equilibrium states in Minkowski space-time do not straightforwardly generalize to curved space-times \cite{RMW, NDB, SAF}.

Magnetic fields play a pivotal role in exploring various astrophysical phenomena, ranging from neutron stars, white dwarfs, pulsars, and black holes to galaxies. Numerous observations underscore the significance of magnetic fields, especially in scenarios where their interaction with general relativity becomes essential. One notable example is their presence in active galactic nuclei \cite{bb1, bb2}, where these nuclei exhibit elevated radiation levels compared to the rest of the galaxy, directly influencing its structure and evolutionary trajectory. Another compelling scenario involves the emergence of relativistic collimated jets within the inner regions of accretion discs, a phenomenon explained by magneto-centrifugal mechanisms \cite{bb3, bb4}. Magnetic fields also play a crucial role in thermal processes within strong magnetic neutron stars \cite{bb5, bb6}, primordial accretion disks \cite{MAL}, molecular cloud formation and evolution \cite{JW}, formation of direct collapse of black holes \cite{MAL2}, population III star formation \cite{CRS}, particle motion and acceleration around a Schwarzschild black hole \cite{AA}, investigation of exact superposition of a central static black hole with a surrounding thin disk \cite{ACGP}, and planetary nebulae and post-AGB nebulae \cite{LS}. Additionally, magnetic fields contribute to various aspects of stellar objects and astronomical scenarios, as detailed in references \cite{WJG, IAG, DG, NSK, LRP, JW2, PS, DGS}.

Analytical models describing astrophysical entities are often constructed based on solutions derived from Einstein's field equations. In the pursuit of more realistic models for compact stellar systems, modifications to the energy-momentum tensor-the source of Einstein's equations are introduced to incorporate additional physical properties, such as electromagnetic fields. Bonnor was the first to consider this approach for deriving such a solution, addressing both radial and longitudinal fields \cite{WBB, WBB2}. Subsequently, Melvin revisited this Bonnor solution, which is now recognized as Bonnor-Melvin magnetic universe \cite{MM}. Numerous authors have attempted to construct exact solutions to the field equations in the presence of magnetic fields, including Manko solution \cite{VM,VM2}, axisymmetric Einstein-Maxwell solutions with a cosmological constant \cite{MA2, MZ, JV}, models of axially symmetric thin disks with finite extension \cite{ECD}, and general magnetostatic axisymmetric exact solutions \cite{CHGD}, among others. In the context of quantum mechanical systems, the Einstein-Maxwell solutions with a cosmological constant \cite{MA2, MZ, JV} has recently attracted attention in research domain. These investigations include quantum motion of scalar particles interacting with linear and Cornell-type potentials \cite{FAAB}, and rainbow gravity effects ts on scalar bosons \cite{FAAB2}. In addition, relativistic wave equation for spin $1/2$ particles in the Melvin space-time was investigated in Ref. \cite{LCNS3}. This emerging body of research significantly contributes to our understanding of the interplay between quantum systems and the geometries of curved space-times.

Our objective is to investigate the relativistic quantum dynamics of quantum oscillator fields within a PDM system in the backdrop of axisymmetric Einstein-Maxwell solution featuring a cosmological constant \cite{MA2, MZ, JV}. The PDM oscillator system is examined by substituting the momentum operator $p_{\mu} \to \Big(p_{\mu}+i\,\eta\,X_{\mu}+i\,\mathcal{F}_{\mu}\Big)$ into the Klein-Gordon equation, where $\mathcal{F}_{\mu}=(0, \mathcal{F}_r, 0, 0)$ with $\mathcal{F}_r=\frac{m'(r)}{4\,m(r)}=\frac{f'(r)}{4\,f(r)}$, and $\eta=m_0\,\omega$ with $\omega$ represents the oscillator frequency. We consider two forms of position-dependent mass $m(r)=m_0\,f(r)$, where scalar multiplier $f(r)$ is given by (i) $f(r)=e^{\frac{1}{2}\,\alpha\,r^2}$ and (ii) $f(r) \propto r^{\beta}$, with $\alpha \geq 0$ and $\beta \geq 0$. We solve the Klein-Gordon equation within the chosen space-time background and obtain the approximate energy eigenvalues for each case. Importantly, we demonstrate that the approximate eigenvalue solutions undergo modifications due to the presence of the geometrical topology and the cosmological constant, deviating from the results obtained in Minkowski flat space. This emphasizes the influence of the considered space-time background on the quantum properties of the oscillator fields, revealing deviations from the outcomes predicted in a flat space-time setting. 

We format this paper as follows: In {\it section 2}, we derive the radial equation of the Klein-Gordon wave equation with a position-dependent mass. The radial equation is solved using two different PDM scalar multiplier through special functions, leading to the energy profile of the oscillator fields. In {\it section 3}, we present and discuss the results obtained. Throughout the paper, we adopt a system of units where $c=1=\hbar$. 

\section{PDM quantum oscillators in Einstein-Maxwell-Lambda space-time }

In this section, the relativistic quantum oscillator described by the Klein-Gordon oscillator within the framework of PDM system in an axisymmetric Einstein-Maxwell solution with a cosmological is investigated. This cosmological model in the cylindrical system is described by the following line-element \cite{MA2, MZ, JV, FAAB, FAAB2}
\begin{equation}
ds^{2}=-dt^{2}+dr^{2}+\sigma^{2}\,\sin^{2}\left(\sqrt{2\Lambda}\,r\right)\,d\varphi^{2}+dz^{2},\label{eq:a1}
\end{equation}
where $\sigma$ is the topological parameter which produces an angular deficit and $\Lambda$ is the cosmological constant. The magnetic field strength associated with this metric is given by $H(r)=\sigma\,\sqrt{\Lambda}\,\sin \left(\sqrt{2\Lambda}\,r\right)$ which acts along the $z$-direction. The electromagnetic field tensor associated with this metric is given by ${\bf F}=H(r)\,dr \wedge d\varphi$. This metric is analogous but not exact form to the cosmic string space-time, $ds^2=-dt^2+dr^2+\alpha^2\,r^2\,d\varphi^2+dz^2$ \cite{WAH,ERFM}. Expressing the space-time (\ref{eq:a1}) in the form $ds^2=g_{\mu\nu}\,dx^{\mu}\,dx^{\nu}$, where $\mu,\nu=0,123$, the covariant metric tensor $g_{\mu\nu}$ and its contravariant form $g^{\mu\nu}$ are given by
\begin{equation}
g_{\mu\nu}=\left(\begin{array}{cccc}
-1 & 0 & 0 & 0\\
0 & 1 & 0 & 0\\
0 & 0 & \sigma^2\,\sin^2\left(\sqrt{2\Lambda}\,r\right) & 0\\
0 & 0 & 0 & 1
\end{array}\right),\quad
g^{\mu\nu}=\left(\begin{array}{cccc}
-1 & 0 & 0 & 0\\
0 & 1 & 0 & 0\\
0 & 0 & \frac{1}{\sigma^2\,\sin^2\left(\sqrt{2\Lambda}\,r\right)} & 0\\
0 & 0 & 0 & 1
\end{array}\right).\label{eq:a2}
\end{equation}
Finally, the determinant of the metric tensor $g_{\mu\nu}$ for the above space-time (\ref{eq:a1}) is given by
\begin{equation}
    \det\,(g_{\mu\nu})=g=-\sigma^2\,\sin^2 (\sqrt{2\Lambda}\,r) \label{eq:a3}
\end{equation}
which vanishes on the symmetry axis $r=0$, thus, representing an example of degenerate metrics. 

The relativistic quantum motions of spin-0 scalar particles is described by the following wave equation \cite{FAAB, FAAB2, WG}
\begin{eqnarray}
    \Big[\frac{1}{\sqrt{-g}}\,\partial_{\mu}\,(\sqrt{-g}\,g^{\mu\nu})\,\partial_{\nu})\Big]\,\Psi=m^2_{0}\,\Psi,\label{eq:a4}
\end{eqnarray}
where $m_0$ is the rest mass of the particles and $g$ is the determinant of the metric tensor $g_{\mu\nu}$.

In quantum system, the Klein-Gordon oscillator \cite{SB} can be studied by replacing the momentum four-vector: $p_{\mu} \to (p_{\mu}+i\,\eta\, X_{\mu})$ \cite{BM, HH, HH2}, where $\eta$ is the oscillator frequency, and the four-vector is defined by $X_{\mu}=(0, r, 0, 0)$. This quantum oscillator field has been analyzed by numerous authors in the backdrop of various curved space-time, such as the G\"{o}del-type space-time of Som-Raychaudhuri metric \cite{ZW}, cosmic strings space-time \cite{LCNS,AHEP}, point-like global monopoles \cite{EAFB, SR}, topologically non-trivial \cite{LCNS2} and trivial space-times \cite{FA}, and in the context of Kaluza-Klein theory \cite{JC}. In this analysis, we introduce position-dependent mass (PDM) into the quantum system described by the Klein-Gordon oscillator. Noted that the concept of position-dependent effective mass (PDM) \cite{aa1} has sparked research interest in both classical and quantum mechanics. Such a PDM concept is, in fact, a metaphoric manifestation of coordinate transformation \cite{aa5}. The coordinate transformation, in effect, changes the form of the canonical momentum in classical and the momentum operator in quantum mechanics \cite{aa7, aa8}. In quantum mechanics, the PDM momentum operator is constructed in Refs. \cite{aa7, aa8, aa9, aa10} given by 
\begin{equation}
 \hat{p}({\bf r}) =-\,\Big(\nabla-\frac{\nabla\,m(r)}{4\,m(r)}\Big)\Longleftrightarrow \hat{p}_{j}({\bf r})=-i\,\Big(\partial_{j}-\frac{\partial_{j}\,m(r)}{4\,m(r)}\Big),\label{om}   
\end{equation}
where $j=1,2,3$. 

Nonetheless, efforts have been made to incorporate PDM settings into Dirac and KG relativistic equations by assuming $m \to m_0 +S(r)=m(r)$, where $m_0$ represents the rest mass energy, $S(r)$ is the Lorentz scalar potential \cite{WG}, and $m(r)$ represents PDM \cite{aa9, aa10}. However, in this present study, we refrain from utilizing this assumption. Instead, we assert that similar to the conventional textbook approach, where the momentum operator $p_{j}=-i\,\partial_{j}$ is employed for constant mass in the relativistic wave equations, an analogous procedure should be adopted for PDM-relativistic quantum particles. Thus, for PDM particles, both relativistic and non-relativistic, the PDM-momentum operator (\ref{om}) should substitute the constant mass textbook momentum operator $p_{j}=-i\,\partial_{j}$. In our methodological approach, we adopt this PDM assumption and explore the impacts of the gravitational field generated by magnetic space-time on certain confined PDM KG-oscillators. Therefore, PDM system Klein-Gordon oscillator can studied by replacing the momentum four-vector as follows \cite{aa7, aa8, aa9, aa10, OM2, OM3}: 
\begin{equation}
    p_{\mu} \to \Big(p_{\mu}+i\,\eta\,X_{\mu}\Big) \to \Big(p_{\mu}+i\,\eta\,X_{\mu}+i\,\mathcal{F}_{\mu}\Big),\label{eq:aa4} 
\end{equation}
where we have defined this new four-vector $\mathcal{F}_{\mu}$ as follows \cite{aa7, aa9, aa10, OM2, OM3}:
\begin{equation}
    \mathcal{F}_{\mu}=(0, \mathcal{F}_r, 0, 0),\quad \mathcal{F}_r=\frac{m'(r)}{4\,m(r)}=\frac{f'(r)}{4\,f(r)},\label{eq:a5}
\end{equation}
where our assumption is that $m(r)=m_0\,f(r)$ depends only on the axial or radial coordinate.

Thus, the relativistic wave equation (\ref{eq:a4}) describing PDM Klein-Gordon oscillator using (\ref{eq:a4}) becomes 
\begin{eqnarray}
\Bigg[\frac{1}{\sqrt{-g}}\,\Big(\partial_{\mu}+\eta\,X_{\mu}+\mathcal{F}_{\mu}\Big)\,\Big\{\sqrt{-g}\,g^{\mu\nu}\,\Big(\partial_{\nu}-\eta\,X_{\nu}-\mathcal{F}_{\nu}\Big)\Big\}\Bigg]\,\Psi=m^2_{0}\,\Psi.\label{eq:a6}
\end{eqnarray}

Expressing this wave equation (\ref{eq:a6}) in the space-time background (\ref{eq:a1}) and using (\ref{eq:a2})--(\ref{eq:a3}), we obtain the following second-order differential equation involving both time and space coordinates as follows:
\begin{eqnarray}
\Bigg[-\frac{d^2}{dt^2}+\frac{d^2}{dr^2}+\frac{\kappa}{\tan (\kappa\,r)}\,\frac{d}{dr}+\mathcal{M} (r)+\frac{1}{\sigma^2\,\sin^2 (\kappa\,r)}\,\frac{d^2}{d\phi^2}+\frac{d^2}{dz^2}-m^2_{0}\Bigg]\,\Psi=0.\label{eq:a7}
\end{eqnarray}
where $\kappa=\sqrt{2\,\Lambda}$ and $\mathcal{M} (r)$ is given by 
\begin{equation}
\mathcal{M} (r)=\frac{3}{16}\,\frac{f'^2}{f^2}-\frac{f''}{4\,f}-\frac{\kappa}{\tan (\kappa\,r)}\,\frac{f'}{4\,f}-\frac{\eta\,\kappa\,r}{\tan (\kappa\,r)}-\eta-\eta^2\,r^2-\eta\,r\,\,\frac{f'}{2\,f}.\label{eq:a8}
\end{equation}

One can see that the second-order differential equation (\ref{eq:a7}) depends only on the radial coordinate $r$ and independent of the coordinates $(t, \varphi, z)$. In mathematical physics, this type of differential equation can easily be solved using the method of separation of variables. In this work, we choose an ansatz for the wave function ($\Psi$) in terms of different variables with the radial function $\psi(r)$ as follows:
\begin{equation}
    \Psi (t, r, \varphi, z) =e^{i\,(-E\,t+\ell\,\phi+k_{z}\,z)}\,\psi(r), \label{eq:a9}
\end{equation}
where $E$ is the particle's energy, $\ell=0,\pm\,1,\pm\,2,....$ are the eigenvalues of the angular quantum number, and $k_z$ is an arbitrary constant. For axisymmetric space-time, we can set $k_z=0$ in the wave function for simplicity.

Thereby, substituting the total wave function (\ref{eq:a9}) into the differential equation (\ref{eq:a7}) results the following a linear, second-order homogeneous differential equation form given by
\begin{equation}
\psi''(r)+\frac{\kappa}{\tan (\kappa\,r)}\,\psi' (r)+\Bigg[E^2-m^2_{0}+\mathcal{M} (r)-\frac{\ell^2}{\sigma^2\,\sin^2 (\kappa\,r)} \Bigg]\,\psi(r)=0,\label{eq:a10}
\end{equation}
where $\mathcal{M} (r)$ is given in Eq. (\ref{eq:a8}).

Now, we solve this second-order radial wave equation (\ref{eq:a10}) using different cases of $f(r)$ other than unity. For that purpose, we choose two such function or multiplier, namely, (i) $f(r)=\exp (\alpha\,r^2/2)$, and (ii) a power law type dimensionless scalar multiplier $f(r) \propto r^{\beta}$, where $\alpha \geq 0, \beta \geq 0$ \cite{aa7, OM2, OM3}. Using these, we solve the radial equation (\ref{eq:a10}) and obtain the energy profiles and the radial wave function in each cases.

\subsection{PDM scalar multiplier: $f(r)=e^{\alpha\,r^2/2}$ } 

To solve the radial equation (\ref{eq:a10}), we choose the following scalar multiplier
\begin{equation}
    f(r)=e^{\alpha\,r^2/2},\label{eq:f1}
\end{equation}
where $\alpha \geq 0$.

Thereby, substituting this $f(r)$ into the differential equation (\ref{eq:a10}), we obtain the following differential equation form:
\begin{eqnarray}
\psi''(r)+\frac{\kappa}{\tan (\kappa\,r)}\psi' (r)+\Bigg[E^2-m^2_{0}-\Big(\frac{\alpha}{4}+\eta\Big)^2r^2-\Big(\frac{\alpha}{4}+\eta\Big)\Big\{1+\frac{\kappa\,r}{\tan (\kappa\,r)}\Big\}-\frac{\ell^2}{\sigma^2\,\sin^2 (\kappa\,r)} \Bigg]\psi(r)=0.\label{eq:a12}
\end{eqnarray}

In order to solve equation (\ref{eq:a12}), one can follows different approximation schemes known in the quantum systems. In this analysis, we choose an approximation where the axial distance $r$ is small such that $\sin r \approx r$ and $\tan r \approx r$. Therefore, using this approximation up to first order, we obtain the following equation from the Eq. (\ref{eq:a12}) given by
\begin{eqnarray}
\psi''(r)+\frac{1}{r}\,\psi' (r)+\Big(\Theta-\omega^2\,r^2-\frac{\iota^2}{r^2} \Big)\,\psi(r)=0,\label{eq:a13}
\end{eqnarray}
where we set
\begin{eqnarray}
    \Theta=E^2-m^2_{0}-2\,\omega,\quad \omega=\frac{\alpha}{4}+\eta,\quad \iota=\frac{|\ell|}{|\sigma|\,\kappa}.\label{eq:a14}
\end{eqnarray}
Noted that one may consider up to the second order approximation which we left for the future investigations.

At this stage, we can consider a substitution via $\psi=\frac{R(r)}{\sqrt{r}}$ into the equation (\ref{eq:a13}) results
\begin{equation}
    \Bigg[\frac{d^2}{dr^2}-\omega^2\,r^2-\frac{(\iota^2-1/4)}{r^2} +\Theta\Bigg]\,R(r)=0.\label{eq:a15}
\end{equation}

This Eq. (\ref{eq:a15}) is the radial equation that describes the position-dependent mass Klein–Gordon oscillator in the context of a magnetic space-time (\ref{eq:a1}) featuring a cosmological constant and a topological parameter. Equation (\ref{eq:a15}) is the Schrodinger-like differential equation form whose solutions are well-known in the literature. Substituting the following solution
\begin{equation}
    R(r)=r^{\iota+1/2}\,e^{-\frac{1}{2}\,\omega\,r^2}\,G(r),\label{eq:a16}
\end{equation}
into the Eq. (\ref{eq:a15}) and performing a transformation to a new variable via $s=\omega\,r^2$, we obtain
\begin{equation}
s\,G''(s)+\Big(1+\iota-s\Big)\,G'(s)-\Bigg(\frac{1}{2}+\frac{\iota}{2}-\frac{\Theta}{4\,\omega}\Bigg)\,G(s)=0.\label{eq:a18}
\end{equation}

Equation (\ref{eq:a18}) is the confluenct hypergeometric second-order differential equation form \cite{MA, GEA} and the solution of this Eq. (\ref{eq:a18}) is given by the confluent hypergeometric function given by 
\begin{equation}
G(s)={}_1 F_1 \Bigg(\frac{1}{2}+\frac{\iota}{2}-\frac{\Theta}{4\,\omega},1+\iota;s\Bigg).\label{eq:a19}
\end{equation}

\begin{figure}
\begin{subfigure}[b]{0.5\textwidth}
\includegraphics[width=2.4in,height=1.7in]{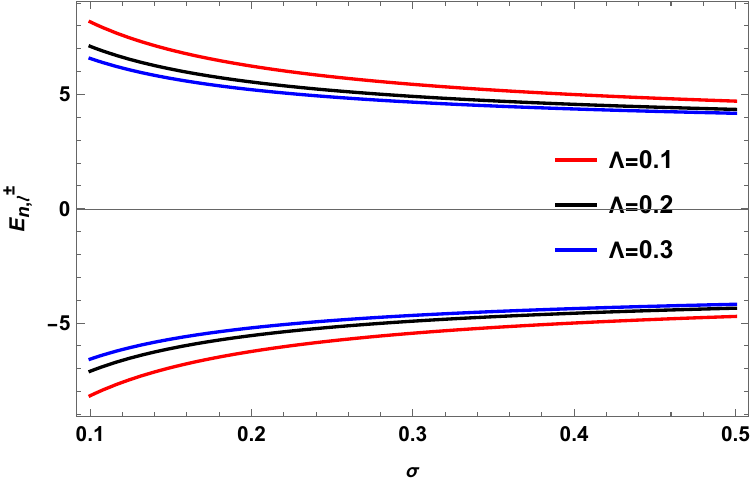}
\caption{$n=1=\eta$}
\label{fig:1 (a)}
\end{subfigure}
\hfill
\begin{subfigure}[b]{0.5\textwidth}
\includegraphics[width=2.4in,height=1.7in]{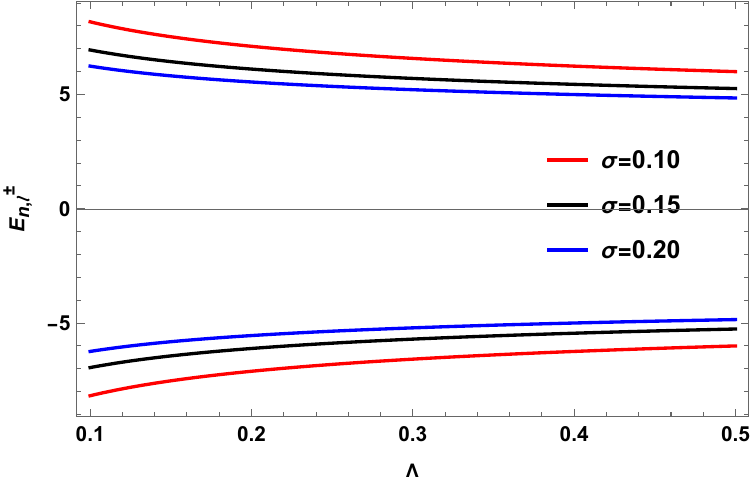}
\caption{$n=1=\eta$}
\label{fig:1 (b)}
\end{subfigure}
\hfill\\
\begin{subfigure}[b]{0.5\textwidth}
\includegraphics[width=2.4in,height=1.7in]{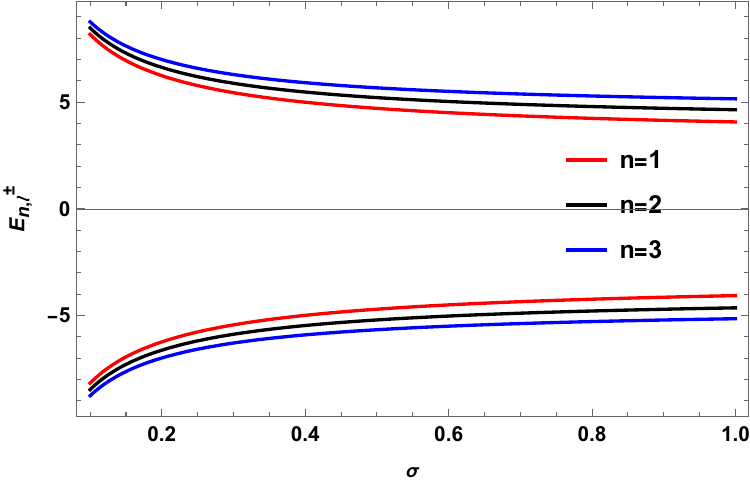}
\caption{$\eta=1$, $\Lambda=0.1$}
\label{fig:1 (c)}
\end{subfigure}
\hfill
\begin{subfigure}[b]{0.5\textwidth}
\includegraphics[width=2.4in,height=1.7in]{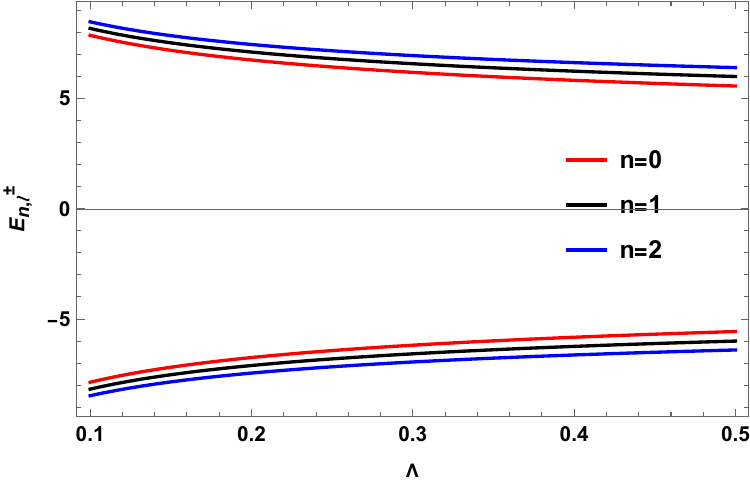}
\caption{$\eta=1$, $\sigma=0.1$}
\label{fig:1 (d)}
\end{subfigure}
\hfill\\
\begin{subfigure}[b]{0.5\textwidth}
\includegraphics[width=2.4in,height=1.7in]{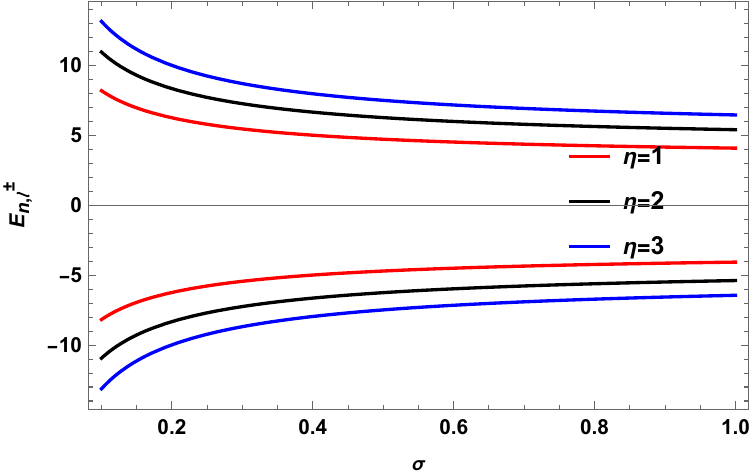}
\caption{$\Lambda=0.1$, $n=1$}
\label{fig:1 (e)}
\end{subfigure}
\hfill
\begin{subfigure}[b]{0.5\textwidth}
\includegraphics[width=2.4in,height=1.7in]{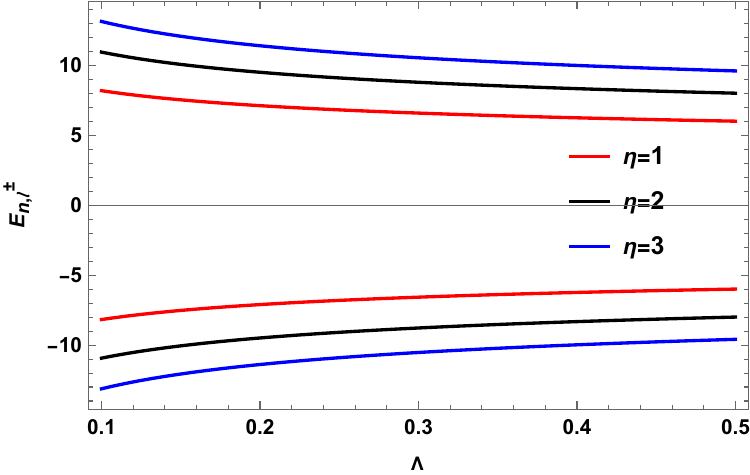}
\caption{$\sigma=0.1$, $n=1$}
\label{fig:1 (f)}
\end{subfigure}
\caption{The energy spectrum with parameter $\sigma$ and cosmological constant $\Lambda$ for various values of other parameters. Here $\ell=1=\alpha=m_0$. }
\label{fig: 1}
\end{figure}

From the asymptotic behavior of the hypergeometric function, it is necessary that this function ${}_1 F_{1}(s)$ when expanding in a power series of must be a finite degree polynomial function of $s$ having degree $n$, and the quantity $\Big(\frac{1}{2}+\frac{\iota}{2}-\frac{\Theta}{4\,\omega}\Big)$ must be a non-positive integer. Therefore, in that scenario, we have the following relation 
\begin{equation}
    \frac{1}{2}+\frac{\iota}{2}-\frac{\Theta}{4\,\omega}=-n \quad (n=0,1,2,3,...).\label{eq:a20}
\end{equation}
Simplification of the above relation results the following approximate energy expression given by
\begin{equation}
    E_{n,\ell}=\pm\,\sqrt{m^2_{0}+(\alpha+4\,\eta)\,\Bigg(n+1+\frac{|\ell|}{2\,\sqrt{2\,\Lambda}\,|\sigma|} \Bigg)}.\label{eq:a21}
\end{equation}

\begin{figure}
\begin{subfigure}[b]{0.5\textwidth}
\includegraphics[width=2.6in,height=1.7in]{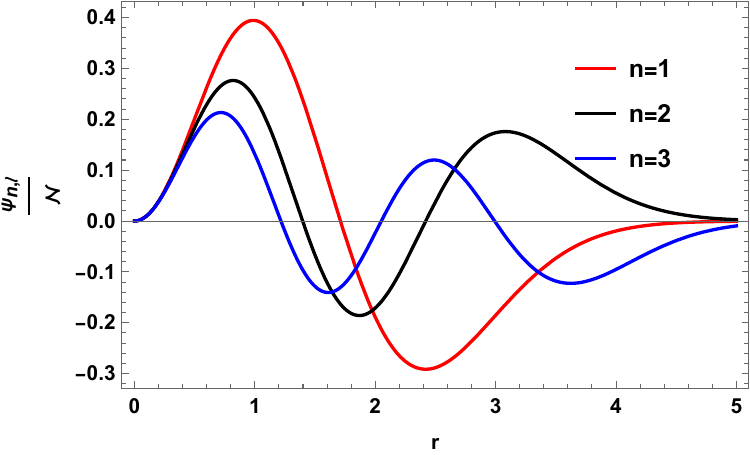}
\caption{$\Lambda=0.5=\sigma, \eta=1$}
\label{fig:2 (a)}
\end{subfigure}
\hfill
\begin{subfigure}[b]{0.5\textwidth}
\includegraphics[width=2.6in,height=1.7in]{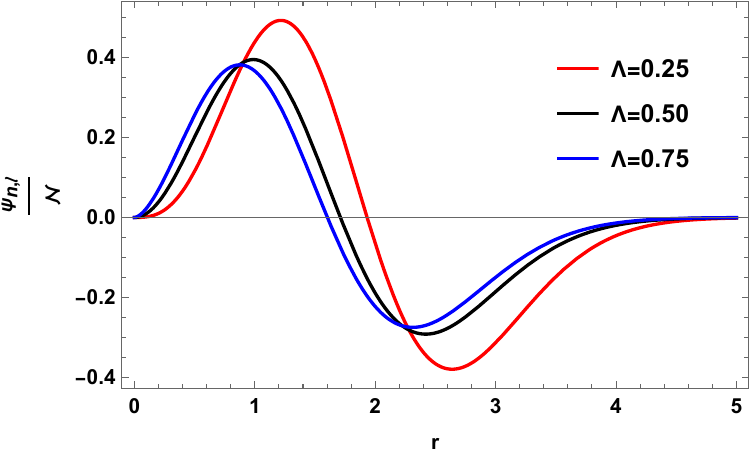}
\caption{$\eta=1=n, \sigma=0.5$}
\label{fig:2 (b)}
\end{subfigure}
\hfill\\
\begin{subfigure}[b]{0.5\textwidth}
\includegraphics[width=2.6in,height=1.7in]{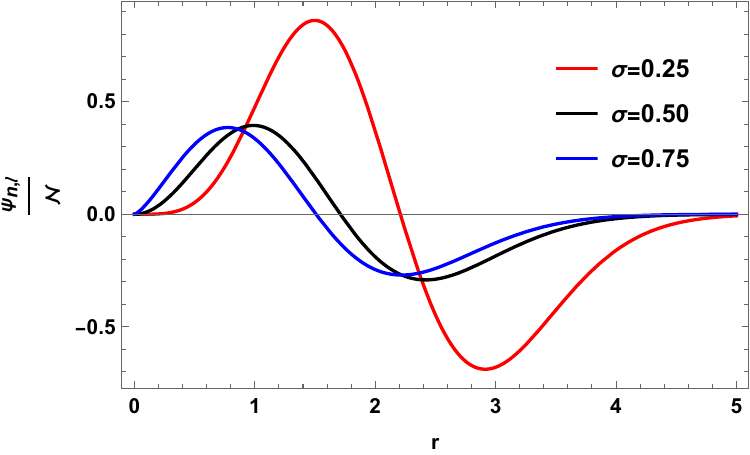}
\caption{$\Lambda=0.5, n=1=\eta$}
\label{fig:2 (c)}
\end{subfigure}
\hfill
\begin{subfigure}[b]{0.5\textwidth}
\includegraphics[width=2.6in,height=1.7in]{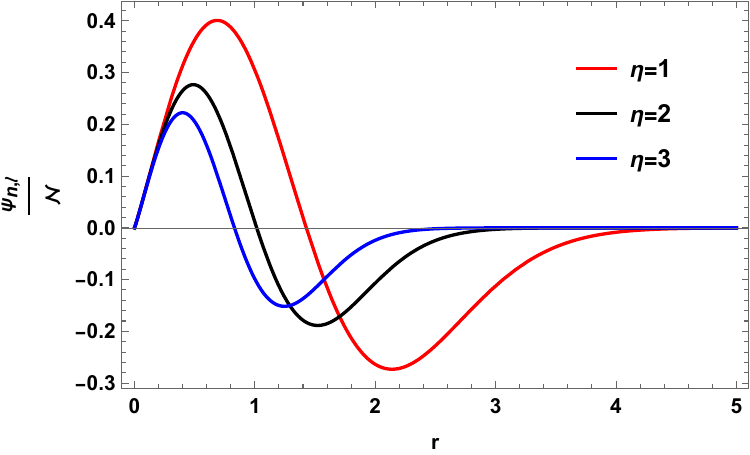}
\caption{$\Lambda=0.75=\sigma, n=1$}
\label{fig:2 (d)}
\end{subfigure}
\hfill\\
\begin{subfigure}[b]{0.5\textwidth}
\includegraphics[width=2.6in,height=1.7in]{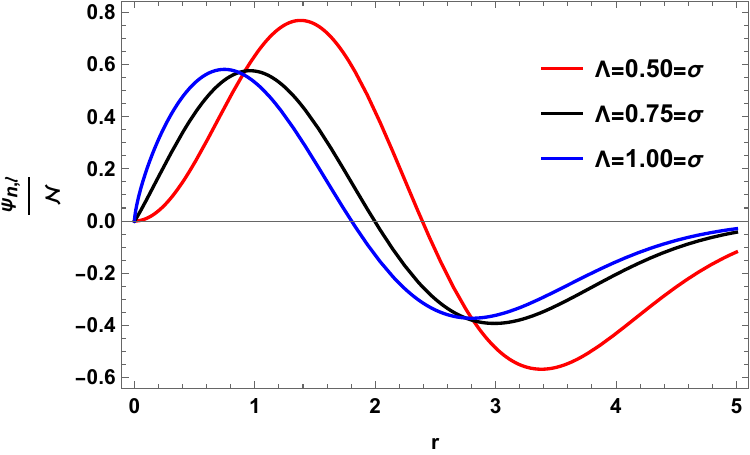}
\caption{$n=1, \eta=0.5$}
\label{fig:2 (e)}
\end{subfigure}
\hfill
\begin{subfigure}[b]{0.5\textwidth}
\includegraphics[width=2.6in,height=1.7in]{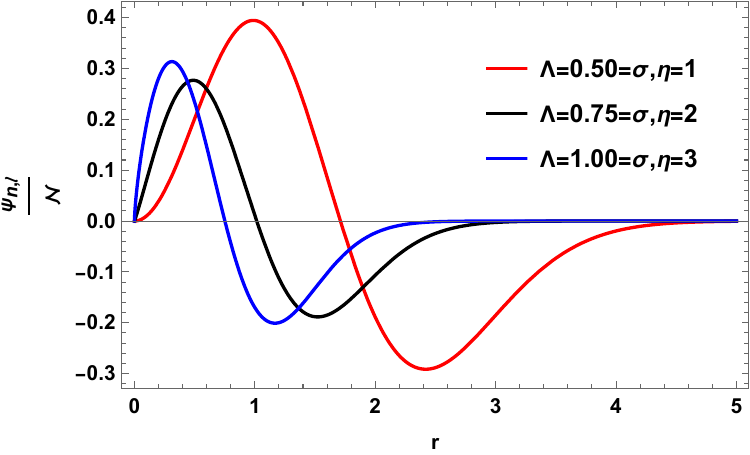}
\caption{$n=1$}
\label{fig:2 (f)}
\end{subfigure}
\caption{The radial wave function for various values of parameters. Here $\ell=1, \alpha=0.1$. }
\label{fig: 2}
\end{figure}

The radial wave function of the quantum oscillator fields are given by
\begin{equation}
    \psi_{n,\ell} (r)=\mathcal{N}\,r^{\frac{|\ell|}{\sqrt{2\,\Lambda}\,|\sigma|}}\,e^{-\frac{1}{2}\,\Big(\frac{\alpha}{4}+\eta\Big)\,r^2}\,{}_1 F_1 \Bigg(-n, 1+\frac{|\ell|}{\sqrt{2\,\Lambda}\,|\sigma|}; \Big(\frac{\alpha}{4}+\eta\Big)\,r^2\Bigg),\label{eq:a22}
\end{equation}
where $\mathcal{N}$ is the normalization constant.

Equation (\ref{eq:a21}) represents the approximate relativistic energy eigenvalue, while equation (\ref{eq:a22}) characterizes the radial wave function of PDM quantum oscillator fields in the context of a curved space-time produced by an axisymmetric Einstein-Maxwell space-time featuring a non-zero cosmological constant. Here, the position-dependent mass (PDM) is considered by $m(r)=m_0\,e^{\alpha\,r^2/2}$. Notably, the approximate energy levels and the radial wave function of oscillator fields is influenced by the magnetic field parameters ($\sigma, \Lambda$), the coefficient parameter $\alpha$ and undergoes modification. Furthermore, the approximate eigenvalue solution undergoes alteration as influenced by the quantum numbers $\{n, \ell\}$. Importantly, the modifications diverge from the outcomes derived in the flat space context for the quantum oscillator field.

Figure 1 shows the approximate energy spectrum, $E^{\pm}_{n,\ell}$ of the quantum oscillator field as defined in Eq. (\ref{eq:a21}), with respect to the geometric topology parameter $\sigma$ and the cosmological constant $\Lambda$, considering varying values of other additional parameters. Notably, an observable trend is the gradual decrease in the energy levels as these additional parameters increase. Subsequently, increasing values of these parameters result in a downward shift in the energy levels as seen in the sub-figures (a)-(b), whereas an upward shift is evident in the sub-figures (c)--(f). In Figure 2, we visualized the behavior of the radial wave function (\ref{eq:a22}) of the quantum oscillator field, and highlighting its response to diverse values of one or more parameters involving in it.

If one choose $\alpha=0$, then the PDM becomes a constant, that is, $m(r)=m_0$. In that case, considering $\eta=m_0\,\Omega$ one will get back the quantum oscillator system known as the Klein-Gordon oscillator. The approximate energy eigenvalue of the relativistic quantum oscillator fields is given by the following expression 
\begin{equation}
    E_{n,\ell}=\pm\,\sqrt{m^2_{0}+4\,m_0\,\Omega\,\Bigg(n+\frac{|\ell|}{2\,\sqrt{2\,\Lambda}\,|\sigma|}+1 \Bigg)}.\label{eq:a23}
\end{equation}
And that the radial wave function is given by
\begin{equation}
    \psi_{n,\ell} (r)=\mathcal{N}\,r^{\frac{|\ell|}{\sqrt{2\,\Lambda}\,|\sigma|}}\,e^{-\frac{1}{2}\,m_0\,\Omega\,r^2}\,{}_1 F_1 \Bigg(-n, 1+\frac{|\ell|}{\sqrt{2\,\Lambda}\,|\sigma|}; m_0\,\Omega\,r^2\Bigg).\label{eq:a24}
\end{equation}

\subsection{PDM scalar multiplier: $f (r)=c\,r^{\beta}$, where $\beta \geq 0$ }

In this part, we solve the radial equation (\ref{eq:a10}) using the following scalar multiplier
\begin{equation}
    f (r)=c\,r^{\beta},\quad \beta \geq 0.\label{fun-2}
\end{equation}
Thereby, substituting this into the equation (\ref{eq:a10}), we obtain the following differential equation
\begin{eqnarray}
&&\psi''(r)+\frac{\kappa}{\tan (\kappa\,r)}\,\psi' (r)+\Bigg[E^2-m^2_{0}-\eta-\frac{1}{2}\,\eta\,\beta+\Big(-\frac{\beta^2}{16}+\frac{\beta}{4}\Big)\,\frac{1}{r^2}-\eta^2\,r^2\nonumber\\
&&-\frac{\kappa\,\beta}{4\,\tan (\kappa\,r)}\,\frac{1}{r}-\frac{\ell^2}{\sigma^2\,\sin^2 (\kappa\,r)}-\eta\,\frac{\kappa\,r}{\tan (\kappa\,r)} \Bigg]\,\psi(r)=0.\label{eq:b1}
\end{eqnarray}

Following the approximation done in the previous case, we obtain a standard second-order differential equation from Eq. (\ref{eq:b1}) given by  
\begin{eqnarray}
\psi''(r)+\frac{1}{r}\,\psi' (r)+\Big(\Pi-\eta^2\,r^2-\frac{\tau^2}{r^2} \Big)\,\psi(r)=0,\label{eq:b2}
\end{eqnarray}
where we set
\begin{equation}
    \Pi=E^2-m^2_{0}-2\,\eta\,-\frac{1}{2}\,\eta\,\beta,\quad \tau=\sqrt{\frac{\beta^2}{16}+\frac{\ell^2}{2\,\Lambda\,\sigma^2}}.\label{eq:b3}
\end{equation}
Here, also performing a transformation to new function via $\psi(r)=\frac{R(r)}{\sqrt{r}}$ into the equation (\ref{eq:b2}) results
\begin{eqnarray}
\Bigg[\frac{d^2}{dr^2}-\eta^2\,r^2-\frac{(\tau^2-1/4)}{r^2} +\Pi\Bigg]\,\psi(r)=0,\label{eq:b4}
\end{eqnarray}
As stated earlier, regular solution is one of the requirement of the quantum systems under investigation. Let a regular solution of the differential Eq. (\ref{eq:b2}) has the following form
\begin{equation}
    R(r)=r^{\tau+1/2}\,e^{-\frac{1}{2}\,\eta\,r^2}\,F(r),\label{eq:b5}
\end{equation}
where $F(r)$ is an unknown function.

Substituting this solution (\ref{eq:b4}) into the equation (\ref{eq:b4}), we obtain the following differential equation form:
\begin{equation}
    F''(r)+\Big(1+2\,\tau-2\,r^2\,\eta \Big)\,F'(r)/r+\Big(\Pi-2\,(1+\tau)\,\eta\Big)\,F(r)=0.\label{eq:b6}
\end{equation}
Finally, performing a transformation to a new variable via $u=\eta\,r^2$ into this equation (\ref{eq:b6}), we obtain a standard second-order differential equation form given by
\begin{equation}
u\,F''(u)+\Big(1+\tau-u\Big)\,F'(u)-\Bigg(\frac{1}{2}+\frac{\tau}{2}-\frac{\Pi}{4\,\eta}\Bigg)\,F(u)=0.\label{eq:b7}
\end{equation}
The above equation is again the confluent hypergeometric second-order differential equation form \cite{MA, GEA} and the solution of this Eq. (\ref{eq:b7}) is given by the confluent hypergeometric function:
\begin{equation}
F(u)={}_1 F_1 \Bigg(\frac{1}{2}+\frac{\tau}{2}-\frac{\Pi}{4\,\eta},1+\tau; u\Bigg).\label{eq:b8}
\end{equation}

\begin{figure}
\begin{subfigure}[b]{0.5\textwidth}
\includegraphics[width=2.4in,height=1.7in]{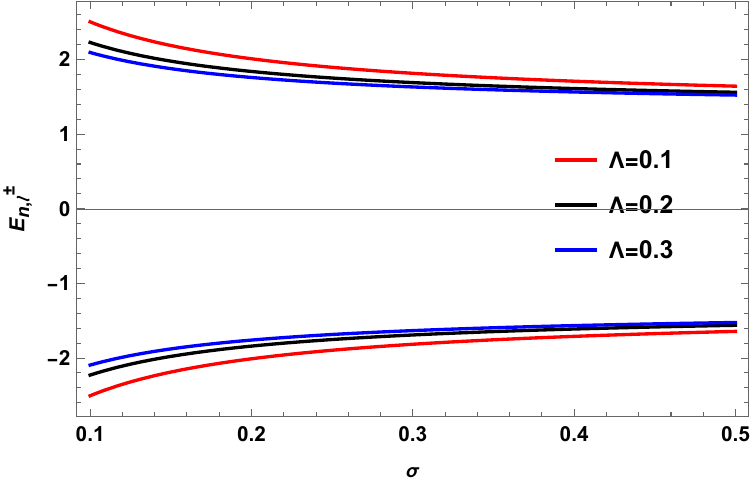}
\caption{$n=1,\eta=0.1$}
\label{fig:3 (a)}
\end{subfigure}
\hfill
\begin{subfigure}[b]{0.5\textwidth}
\includegraphics[width=2.4in,height=1.7in]{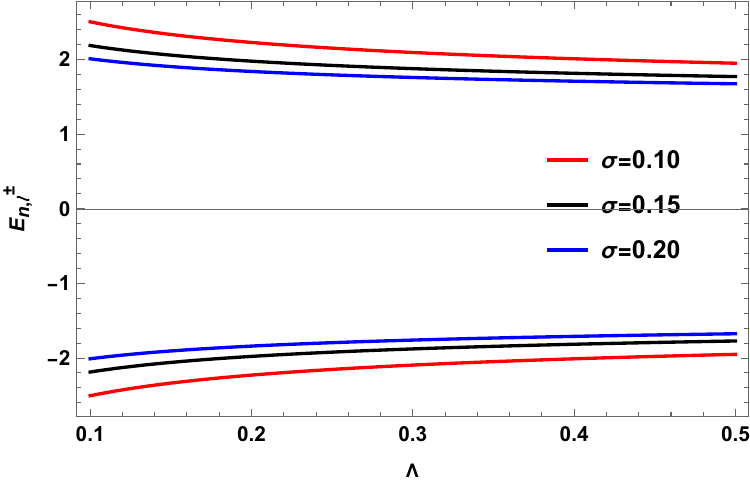}
\caption{$n=1,\eta=0.1$}
\label{fig:3 (b)}
\end{subfigure}
\hfill\\
\begin{subfigure}[b]{0.5\textwidth}
\includegraphics[width=2.4in,height=1.7in]{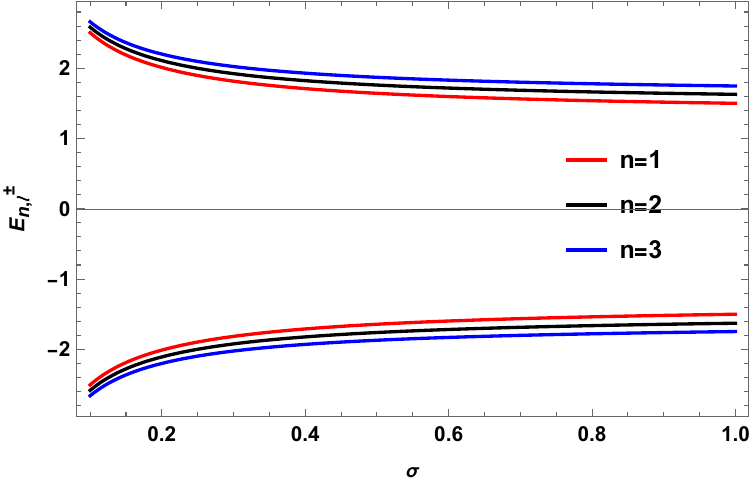}
\caption{$\eta=0.1=\Lambda$}
\label{fig:3 (c)}
\end{subfigure}
\hfill
\begin{subfigure}[b]{0.5\textwidth}
\includegraphics[width=2.4in,height=1.7in]{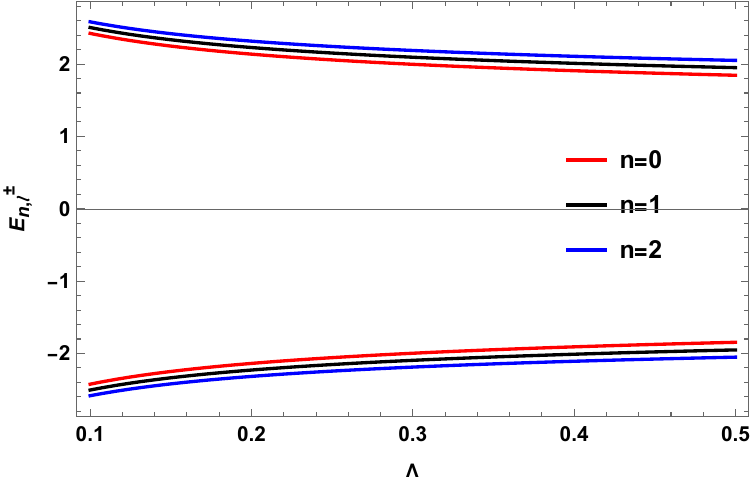}
\caption{$\eta=0.1=\sigma$}
\label{fig:3 (d)}
\end{subfigure}
\hfill\\
\begin{subfigure}[b]{0.5\textwidth}
\includegraphics[width=2.4in,height=1.7in]{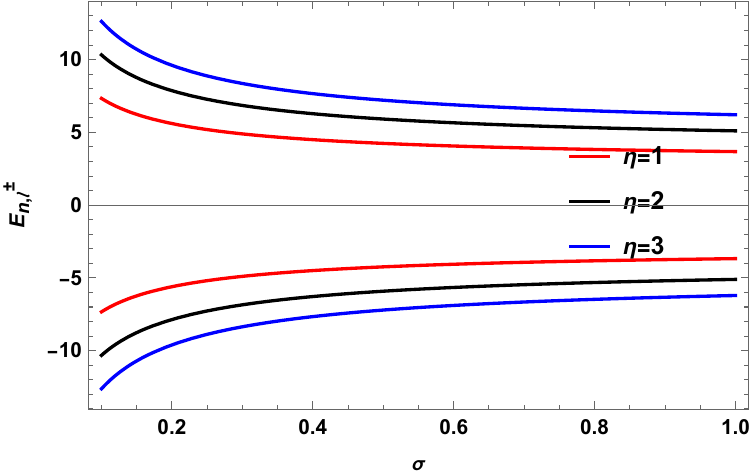}
\caption{$\Lambda=0.1$, $n=1$}
\label{fig:3 (e)}
\end{subfigure}
\hfill
\begin{subfigure}[b]{0.5\textwidth}
\includegraphics[width=2.4in,height=1.7in]{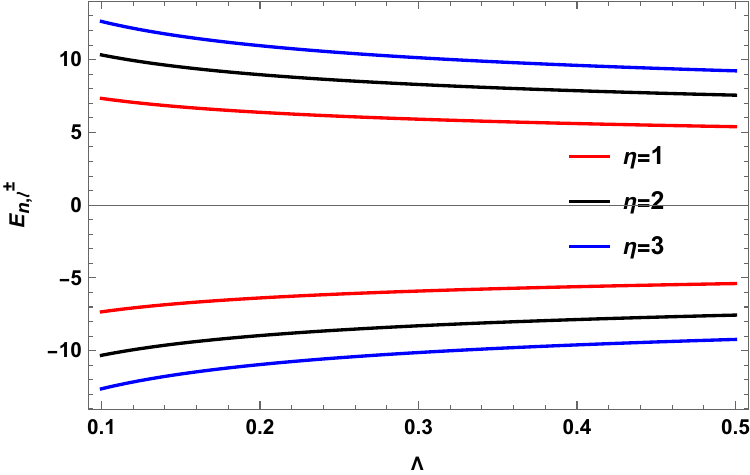}
\caption{$\sigma=0.1$, $n=1$}
\label{fig:3 (f)}
\end{subfigure}
\caption{The energy spectrum with parameter $\sigma$ and cosmological constant $\Lambda$ for various values of other parameters. Here $\ell=1=m_0, \beta=0.1$. }
\label{fig: 3}
\end{figure}

Applying similar argument on the confluent hypergeometric function as done in the previous case, this function is a finite degree polynomial function of $u$ having degree $n$ provided we have the following relation
\begin{equation}
    \frac{1}{2}+\frac{\tau}{2}-\frac{\Pi}{4\,\eta}=-n \quad (n=0,1,2,3,...).\label{eq:b9}
\end{equation}
Simplification of the above relation results the following approximate energy expression
\begin{equation}
    E_{n,\ell}=\pm\,\sqrt{m^2_{0}+4\,\eta\,\Bigg(n+\frac{1}{2}\sqrt{\frac{\beta^2}{16}+\frac{\ell^2}{2\,\Lambda\,\sigma^2}}+1+\frac{\beta}{8}\Bigg)}\,.\label{eq:b10}
\end{equation}

\begin{figure}
\begin{subfigure}[b]{0.5\textwidth}
\includegraphics[width=2.6in,height=1.6in]{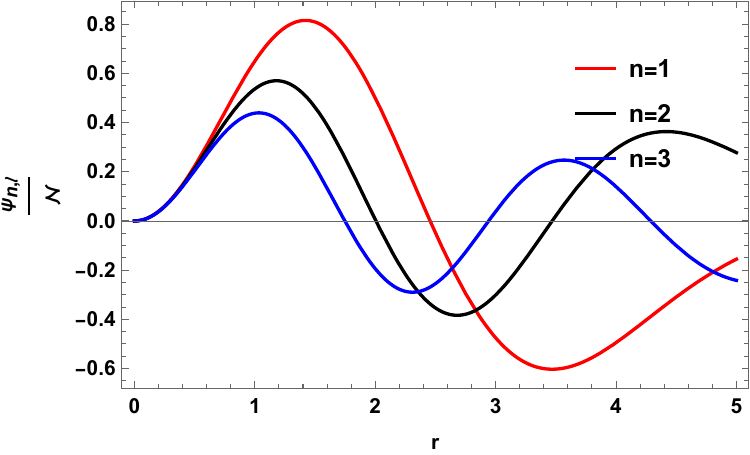}
\caption{$\Lambda=0.5=\sigma, \eta=0.5$}
\label{fig:4 (a)}
\end{subfigure}
\hfill
\begin{subfigure}[b]{0.5\textwidth}
\includegraphics[width=2.6in,height=1.6in]{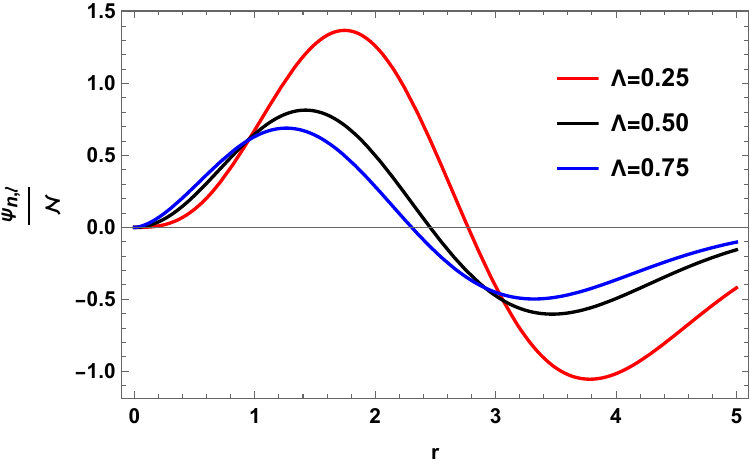}
\caption{$n=1, \sigma=0.5=\eta$}
\label{fig:4 (b)}
\end{subfigure}
\hfill\\
\begin{subfigure}[b]{0.5\textwidth}
\includegraphics[width=2.6in,height=1.6in]{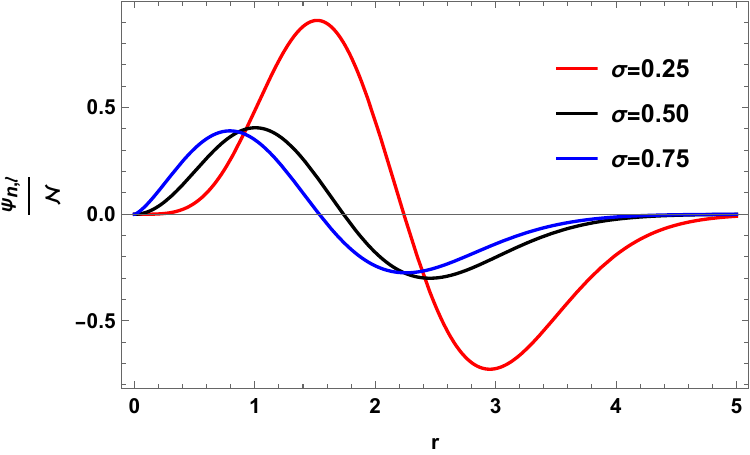}
\caption{$\Lambda=0.5, n=1=\eta$}
\label{fig:4 (c)}
\end{subfigure}
\hfill
\begin{subfigure}[b]{0.5\textwidth}
\includegraphics[width=2.6in,height=1.6in]{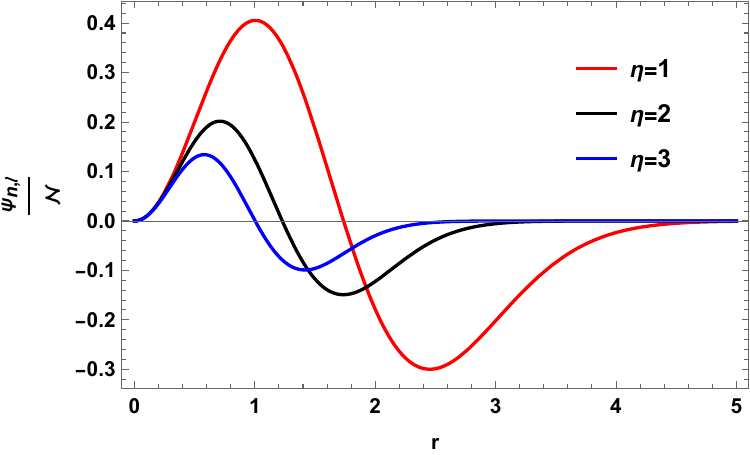}
\caption{$\Lambda=0.5=\sigma, n=1$}
\label{fig:4 (d)}
\end{subfigure}
\caption{The radial wave function for various values of parameters. Here $\ell=1=\beta$. }
\label{fig: 4}
\end{figure}

The radial wave functions are given by
\begin{equation}
    \psi_{n,\ell} (r)=\mathcal{P}\,r^{\sqrt{\frac{\beta^2}{16}+\frac{\ell^2}{2\,\Lambda\,\sigma^2}}}\,e^{-\frac{1}{2}\,\eta\,r^2}\,{}_1 F_1 \Bigg(-n, 1+\sqrt{\frac{\beta^2}{16}+\frac{\ell^2}{2\,\Lambda\,\sigma^2}}; \eta\,r^2\Bigg),\label{eq:b11}
\end{equation}
where $\mathcal{P}$ is the normalization constant.

Equation (\ref{eq:b10}) represents the approximate relativistic energy eigenvalue and Eq. (\ref{eq:b11}) corresponds to the radial wave function of quantum oscillator fields in the background of a magnetic space-time within the framework of position-dependent mass (PDM) characterized by $f(r) \propto r^{\beta}$. It is evident that the approximate energy levels are subject to the influence of geometrical parameters ($\sigma, \Lambda$), and the parameter $\beta$, and this approximate eigenvalue solution undergoes modification by the quantum numbers $\{n, \ell\}$. This modification stand in contrast to the outcomes obtained in a flat space context, emphasizing the impact of both the cosmological background and the position-dependent mass on the approximate eigenvalue solution of the quantum oscillator system.

Figure 3 shows the energy spectrum, $E^{\pm}_{n,\ell}$ of the quantum oscillator field as defined in Eq. (\ref{eq:b10}), with respect to the parameter $\sigma$ and the cosmological constant $\Lambda$, considering varying values of additional parameters. We see that energy level trend is gradually decrease as additional parameters increase. Subsequently, increasing values to these parameters result in a downward shift in the energy levels in sub-figures (a)-(b), whereas an upward shift is evident in sub-figures (c)--(f). In Figure 4, we plot the radial wave function (\ref{eq:b11}) of the oscillator field for diverse values of the parameters involved in it.

In this analysis, one can see that for $\beta=0$ and $\eta=m_0\,\Omega$, the quantum system under investigation reduces to the Klein-Gordon oscillator with constant mass. Therefore, the approximate energy eigenvalue and the radial wave function of the relativistic quantum oscillator field in the background of axisymmetric Einstein-Maxwell space-time with a cosmological constant is given by
\begin{equation}
    E_{n,\ell}=\pm\,\sqrt{m^2_{0}+4\,m_0\,\Omega\,\Bigg(n+\frac{|\ell|}{2\,\sqrt{2\,\Lambda}\,\sigma}+1\Bigg)}\,.\label{eq:b12}
\end{equation}
And
\begin{equation}
    \psi_{n,\ell} (r)=\mathcal{P}\,r^{\frac{|\ell|}{\sqrt{2\,\Lambda}\,\sigma}}\,e^{-\frac{1}{2}\,m_0\,\Omega\,r^2}\,{}_1 F_1 \Bigg(-n, 1+\frac{|\ell|}{\sqrt{2\,\Lambda}\,\sigma}; m_0\,\Omega\,r^2\Bigg),\label{eq:b13}
\end{equation}

The above approximate eigenvalue solution Eqs. (\ref{eq:b12})--(\ref{eq:b13}) is similar to those result obtained earlier in this paper given by Eqs. (\ref{eq:a23})--(\ref{eq:a24}).   

Finally, we want to highlight that in sub-section 2.1, we have chosen the scalar multiplier $f(r)=e^{\frac{1}{2}\,r^{\alpha}}$, where $\alpha>0$, so that PDM becomes $m(r)=m_0\,e^{\frac{1}{2}\,r^{\alpha}}$. While in sub-section 2.2, we have chosen $f(r)=c\,r^{\beta}$, where $\beta>0$ and $c$ is a constant, so that PDM becomes $m(r)=m_0\,c\,r^{\beta}$. 

If one sets $\alpha=0=\beta$, then PDM in both sections becomes $m(r)=m_0$, a constant (by choosing $c=1$ for simplicity). In that scenario, the approximate solutions obtained in Eqs. (\ref{eq:a23})--(\ref{eq:a24}) and Eqs. (\ref{eq:b12})--(\ref{eq:b13}) are similar, otherwise not for $\alpha\neq 0$ and $\beta \neq 0$. Therefore, the results presented in both sub-sections are obviously different.

\section{Conclusions}

In this study, we focused on a specific Einstein-Maxwell solution characterized by a magnetic field and a non-zero positive cosmological constant. Within this magnetic space-time background, we delved into the relativistic dynamics of quantum oscillator fields within the framework of position-dependent mass systems. This position-dependent mass (PDM) system was examined by substituting the momentum four-vector into the Klein-Gordon oscillator equation. We derived the radial equation for the PDM system of the Klein-Gordon oscillator in the presence of this magnetic solution.

Two distinct scalar multipliers, denoted as $f(r)$, were considered, and we determined the approximate relativistic energy eigenvalues and the radial wave function of the system. Significantly, our analysis revealed that various parameters associated with the magnetic fields, such as the geometry's topology ($\sigma$) and the positive cosmological constant ($\Lambda>0$), exerted influence on these eigenvalue solutions, thereby modifying the results compared to those in flat space. To illustrate these findings, we generated several figures depicting the energy spectrum and the radial wave function for various values of the parameters ($\sigma, \Lambda$) involved, as well as the quantum numbers ${n, \ell}$. These figures provided insights into the behavior of the approximate energy spectrum and the wave function as the values of these parameters increased. This comprehensive exploration enhances our understanding of how the interplay between magnetic fields, space-time geometry, and cosmological constants impacts the quantum properties of oscillator fields.

The investigation into position-dependent mass quantum oscillator fields within the framework of Einstein-Maxwell-Lambda space-time has provided us valuable insights into the intricate interplay between quantum mechanics and the gravitational effects. The analysis of such systems not only deepens our understanding of the fundamental principles governing particle dynamics but also sheds light on the behavior of quantum fields in curved space-time backgrounds. The results obtained in this study contribute to the broader exploration of quantum phenomena in diverse gravitational contexts, offering a stepping stone for further research into the rich interconnections between quantum theory and general relativity (GR).

In exploring the interplay between curved space-time and quantum particles, crucial questions arise, prompting further investigation. How does space-time curvature influence particle behavior at the quantum level, and are there distinct quantum signatures in regions of intense gravitational curvature? Venturing into realms with pronounced gravitational effects raises inquiries about its implications for particle dynamics and the nature of quantum fields. Unraveling these mysteries requires bridging conceptual gaps between general relativity and quantum mechanics. Future studies may decipher quantum entanglement in curved space-time or investigate quantum coherence in gravitational fields, guiding us toward a deeper understanding of the quantum-gravitational interface and setting the stage for groundbreaking discoveries reshaping our fundamental cosmos comprehension.

\section*{Data Availability Statement}

No data were generated or analysed in this study.

\section*{Conflict of Interest}

Author declare(s) no such conflict of interests.

\section*{Acknowledgements}

We gratefully acknowledge the anonymous referee's for their invaluable comments and helpful suggestions. F.A. acknowledges the Inter University Centre for Astronomy and Astrophysics (IUCAA), Pune, India, for the granted visiting associateship.


\begin{thebibliography}{1}

\bibitem{M0} S. Hollands, and R. M. Wald, Phys. Reps. {\bf 574}, 1 (2015). 

\bibitem{M1} B. S. DeWitt, Phys. Reps. {\bf 19}, 295 (1975).  

\bibitem{M2} L. H. Ford, {\tt Quantum Field Theory in Curved Spacetime}, arXiv:gr-qc/9707062.

\bibitem{RMW} R. M. Wald, {\tt Quantum Field Theory in Curved Spacetime and Black Hole Thermodynamics}, University of Chicago Press, Chicago (1994).

\bibitem{NDB} N. D. Birrell, and P. C. W. Davies, {\tt Quantum Fields in Curved Space}, Cambridge University Press, Cambridge (1984).

\bibitem{SAF} S. A. Fulling, {\tt Aspects of Quantum Field Theory in Curved Spacetime}, Cambridge University Press, Cambridge (1989).

\bibitem{M3} S. Coleman, and F. De Luccia, Phys. Rev. \textbf{D 21}, 3305 (1980).

\bibitem{M4} A. Vilenkin, and L. H. Ford, Phys. Rev. \textbf{D 26}, 1231 (1982).

\bibitem{M5} D. Harari, and L. Carlos Lousto, Phys. Rev. \textbf{D 42}, 2626 (1990).

\bibitem{M6} M. K. Parikh, and F. Wilczek, Phys. Rev. Lett. \textbf{85}, 5042 (2000).

\bibitem{M7} S. W. Hawking, J. Math. Phys. \textbf{9}, 598 (1968).

\bibitem{M8} S. W. Hawking, Phys. Rev. \textbf{D 14}, 2460 (1976).

\bibitem{M9} S. W. Hawking, Nature \textbf{248}, 30 (1974).

\bibitem{M10} S. W. Hawking, Commun. Math. Phys. \textbf{43}, 199 (1975).

\bibitem{M11} S. W. Hawking, Phys. Rev. \textbf{D 13}, 191 (1976).

\bibitem{M12} G. W. Gibbons, and S. W. Hawking, Phys. Rev. \textbf{D 15}, 2738 (1977).
 
\bibitem{LP} L. Parker, Phys. Rev. {\bf 183}, 1057 (1969).

\bibitem{SWH} S. W. Hawking, Comm. Math. Phys. {\bf 43}, 199 (1975).

\bibitem{WGU} W. G. Unruh, Phys. Rev. {\bf D 14}, 870 (1976).

\bibitem{SW} S. Weinberg, {\tt Cosmology}, Oxford University Press, Oxford (2008). 

\bibitem{bb1} A. P. Lobanov, Astron. Astrophys. {\bf 79}, 330 (1998).

\bibitem{bb2} F. A. Aharonian, MNRAS {\bf 332}, 215 (2002).

\bibitem{bb3} Y. T. Liu, S. L. Shapiro, and B. C. Stephens, Phys. Rev. {\bf D 76}, 084017 (2007).

\bibitem{bb4} W. H. T. Vlemmings, P. J. Diamond and H. Imai, Nature {\bf 440}, 58 (2006).

\bibitem{bb5} D. N. Aguilera, J. A. Pons J A and J. A, Miralles, ApJ {\bf 673} (2008).

\bibitem{bb6} P. Hennebelle, and S. Fromang, Astron. Astrophys. {\bf 477}, 9 (2008).

\bibitem{MAL} M. A. Latif, and D. R. G. Schleicher, Astron. \& Astroph. {\bf 585}, A151 (2016).

\bibitem{JW} J. Wurster, and Z. Y. Li, Front. Astron. Space Sci. {\bf 6},  1 (2019).

\bibitem{MAL2} M. A. Latif, D. R. G. Schleicher and S. Khochfar, ApJ {\bf 945}, 137 (2023).

\bibitem{CRS} C. R. Saad, V. Bromm, M. El. Eid, MNRAS {\bf 516}, 3130 (2022).

\bibitem{AA} A. Abdujabbarov, B. Ahmedov, O. Rahimov and U. Salikhbaev, Phys. Scr. {\bf 89}, 084008 (2014).

\bibitem{ACGP} A. C. G. Pineres, G. G. Reyes and G. A. Gonzalez, Int. J. Mod. Phys. {\bf D 23}, 1450010 (2014).

\bibitem{LS} L. Sabin, A. A. Zijlstra and J. S. Greaves, MNRAS {\bf 376}, 378 (2007).

\bibitem{WJG} W. J. Gray, C. F. McKee, R. I. Klein, MNRAS {\bf 473}, 2124 (2018).

\bibitem{IAG} I. A. Gerrard, C. Federrath, R. Kuruwita, MNRAS {\bf 485}, 5532 (2019).

\bibitem{DG} D. Guszejnov, M. Y. Grudić, P. F. Hopkins, S. S. R. Offner, C.-A. F. Giguère, MNRAS {\bf 496}, 5072 (2020).

\bibitem{NSK} N. S. Kargaltseva, S. A. Khaibrakhmanov, A. E. Dudorov, S. N. Zamozdra and A. G. Zhilkin, Open Astronomy {\bf 31}, 172 (2022).

\bibitem{LRP} L. R. Prole, P. C. Clark, R. S. Klessen, S. C. O. Glover and R. Pakmor, MNRAS {\bf 516}, 2223 (2022).

\bibitem{JW2} J. Wurster, M. R. Bate, D. J. Price and I. A. Bonnell, MNRAS {\bf 511}, 746 (2022).

\bibitem{PS} P. Saha, A. Soam, T. Baug, M. Gopinathan, S. Mondal, T. Ghosh, MNRAS {\bf 513}, 2039 (2022).

\bibitem{DGS} D. García-Senz, R Wissing, R M Cabezón, E Vurgun, M Linares, MNRAS {\bf 518}, 4115 (2023).

\bibitem{WBB} W. B. Bonnor, Proc. Phys. Soc. {\bf A 66}, 145 (1953).

\bibitem{WBB2} W. B. Bonnor, Proc. Phys. Soc. {\bf A 67}, 225 (1954).

\bibitem{MM} M. Melvin, Phys. Lett. {\bf 8}, 65 (1964).

\bibitem{VM} T. Gutsunaev, and V. Manko, Phys. Lett. {\bf A 123}, 215 (1987).

\bibitem{VM2} T. Gutsunaev, and V. Manko, Phys. Lett. {\bf A 132}, 85 (1988).

\bibitem{MA2} M. Astorino, JHEP 06 ({\bf 2012}) 086.

\bibitem{MZ} M. Žofka, Phys. Rev. {\bf D 99}, 044058 (2019).

\bibitem{JV} J. Vesely, and M. \u{Z}ofka, Phys. Rev. {\bf D 100}, 044059 (2019).

\bibitem{ECD} E. Cardona-Rueda, and G. García-Reyes, Indian-J. Phys {\bf 90}, 495 (2016).

\bibitem{CHGD} C. H. García-Duque, and G. García-Reyes, Gen. Relativ. Gravit. {\bf 43}, 3001 (2011).

\bibitem{FAAB} F. Ahmed, and A. Bouzenada, Commun. Theor. Phys. {\bf 76}, 045401 (2024). 

\bibitem{FAAB2} F. Ahmed, and A. Bouzenada, Nucl. Phys. B {\bf 1000}, 116490 (2024).

%\bibitem{FAAB3} F. Ahmed, and A. Bouzenada, arXiv: 2312.06612 [gr-qc].

\bibitem{LCNS3} L. C. N. Santos, and C. C. Barros Jr., Eur. Phys. J. C {\bf 76}, 560 (2016).

\bibitem{WAH} W. A. Hiscock, Phys. Rev. {\bf D 31}, 3288 (1985).

\bibitem{ERFM} E. R. F. Medeiros, and E. R. B. de Mello, Eur. Phys. J. C {\bf 72}, 2051 (2012).

\bibitem{WG} W. Greiner, {\it Relativistic Quantum Mechanics. Wave Equations}, Springer Berlin, Heidelberg (2013).

\bibitem{SB} S. Bruce, and P. Minning, Nuov Cim A {\bf 106}, 711 (1993).

\bibitem{BM} B. Mirza, and M. Mohadesi, Commun. Theor. Phys. {\bf 42}, 664 (2004).

\bibitem{HH} Marc de Montigny, H. Hassanabadi, J. Pinfold and S. Zare, Eur. Phys. J. Plus {\bf 136}, 788 (2021). 

\bibitem{HH2} Marc de Montigny, J. Pinfold, S. Zare and H. Hassanabadi, Eur. Phys. J. Plus {\bf 137}, 54 (2022). 

\bibitem{ZW} Z. Wang, Z. Long, C. Long, and M. Wu, Eur. Phys. J Plus {\bf 130}, 36 (2015).

\bibitem{LCNS} L. C. N. Santos, and C. C. Barros Jr., Eur. Phys. J. C {\bf 78}, 13 (2018).

\bibitem{AHEP} F. Ahmed, Adv. High Energy Phys. {\bf 2020}, 5691025 (2020).

\bibitem{EAFB} E. A. F. Bragança, R. L. L. Vitória, H. Belich, E. R. Bezerra de Mello, Eur. Phys. J. C {\bf 80}, 206 (2020).

\bibitem{SR} F. Ahmed, Sci. Rep. {\bf 12}, 8794 (2022).

\bibitem{LCNS2} L. C. N. Santos, C. E. Mota, and C. C. Barros Jr., Adv. High Energy Phys. {\bf 2019}, 2729352 (2019).

\bibitem{FA} F. Ahmed, Eur. Phys. J. C {\bf 78}, 598 (2018).

\bibitem{JC} J. Carvalho, A. M. de M. Carvalho, E. Cavalcante and C. Furtado, Eur. Phys. J. C {\bf 76}, 365 (2016).

\bibitem{aa1} P. M. Mathews, M. Lakshmanan, Quart. Appl. Math. {\bf 32}, 215 (1974).

\bibitem{aa5} A. Khlevniuk, V. Tymchyshyn, J. Math. Phys. {\bf 59}, 082901 (2018).

\bibitem{aa7} O. Mustafa, Z. Algadhi, Eur. Phys. J. Plus {\bf 134}, 228 (2019).

\bibitem{aa8} M. A. F. dos Santos, I. S. Gomez, B. G. da Costa, O. Mustafa, Eur. Phys. J. Plus {\bf 136}, 96 (2021).

\bibitem{aa9} O. Mustafa, S. H. Mazharimousavi, J. Phys. A: Math. Theor. {\bf 40}, 863 (2007). 

\bibitem{aa10} O. Mustafa, S. H. Mazharimousavi, Int. J. Theor. Phys. {\bf 47}, 1112 (2008).

\bibitem{OM2} O. Mustafa, Ann. Phys. (NY) {\bf 440}, 168857 (2022).

\bibitem{OM3} O. Mustafa, Ann. Phys. (NY) {\bf 446}, 169124 (2022).

\bibitem{MA} M. Abramowitz, and I. A. Stegun, {\tt Handbook of Mathematical Functions with Formulas, Graphs, and Mathematical Tables}, New York: Dover (1972).

\bibitem{GEA} G. E. Andrews, R. Askey and R. Roy, {\tt Special Functions}, Cambridge University Press, Cambridge (1999). 

\end{thebibliography}
\end{document}